\documentclass[conference]{IEEEtran}
\usepackage{algorithm}
\usepackage{algpseudocode}
\usepackage[misc]{ifsym}
\usepackage{bm}
\usepackage{multirow}
\usepackage{amsthm}
\usepackage{amsfonts}
\usepackage{graphicx}
\usepackage{bm}
\usepackage{epstopdf}
\usepackage{epsfig}
\usepackage{soul}
\usepackage[misc]{ifsym}
\usepackage{amsmath}

\IEEEoverridecommandlockouts
\ifCLASSINFOpdf
\else
\fi
\begin{document}
\title{Optimal Design of Compact Receive Array in Industrial Wireless Sensor Networks }

\author{\IEEEauthorblockN{Liangtian Wan, Guangjie Han and Jinfang Jiang}
\IEEEauthorblockA{Department of Information and \\Communication Systems, Hohai University\\
Changzhou, 213022, China\\
Email: wanliangtian1@163.com; hanguangjie@gmail.com;\\
jiangjinfang1989@gmail.com}
\and
\IEEEauthorblockN{Lei Shu}
\IEEEauthorblockA{Guangdong Petrochemical Equipment\\ Fault Diagnosis Key Laboratory,\\
Guangdong University of Petrochemical Technology\\
Guangdong, 525000, China\\
Email: lei.shu@ieee.org}}


%


\maketitle

\begin{abstract}
With the development of wireless communication, industrial wireless sensor networks (IWSNs) plays an important role in monitoring and control systems. In this paper, we extend the application of IWSNs into High Frequency Surface-Wave Radar (HFSWR) system. The traditional antenna is replaced by mobile IWSNs. In combination of the application precondition of super-directivity in HF band and circular topology of IWSNs, a super-directivity synthesis method is presented for designing super-directivity array. In this method, the dominance of external noise is ensured by constraining the Ratio of External to Internal Noise (REIN) of the array, and the desired side lobe level is achieved by implementing linear constraint. By using this method, the highest directivity will be achieved in certain conditions. Using the designed super directive circular array as sub-arrays, the compact receive antenna array is constructed, the purpose of miniaturization is achieved. Simulation verifies that the proposed method is correct and effective, the validity of the proposed method has been proved.
\end{abstract}

\IEEEpeerreviewmaketitle

\section{Introduction}
Given the increasing age of many industrial systems and the dynamic industrial manufacturing market, intelligent and low-cost industrial automation systems are required to improve the productivity and efficiency of such systems [1]. The collaborative nature of industrial wireless sensor networks (IWSNs) brings several advantages over traditional wired industrial monitoring and control systems, including self-organization, rapid deployment, flexibility, and inherent intelligent-processing capability. In this regard, IWSN plays a vital role in creating a highly reliable and self-healing industrial system that rapidly responds to real-time events with appropriate actions [2].

Recently, more attentions have been paid to the research and development of High Frequency Surface-Wave Radar (HFSWR) due to its potentiality in military and civil application. As the working frequency is relatively low ($3 - 30$MHz), the wavelength is extremely long. In order to maintain the performance of radar system, such as signal-to-noise ratio (SNR) and angular resolution, the size of receiving antenna array is always very large. Based on traditional method of array design, the whole received system would occupy the area about ten thousand square meters. The RACE radar of Cananda is composed of 40 monopole wideband elements, the length of the received array is 870m. The length of the received array of the HFSWR designed by Harbin Institute of Technology is 500m, and it occupies 15 thousands square meter. There are many drawbacks in military application due to the extremely large array. First, it is easy to exposure the radar, which can be detected by enemy. Second, the array is not flexible, thus it is hard to manage and maintain. For example, if this system is installed on the surface of warship, the transmitted and received arrays are difficult to design, since the available space of the warship is limited, especially some requirements have to be satisfied. Third, the fisheries and tourism surrounding the coastal area is very developed, if the received array is too large, it would affect the utilization of coastal resource.

The large size would bring lots of problems and trouble in engineering practice. The development of maritime strategic weapons is more and more fast, thus the survivability of HFSWR has to be improved. In other words, it should have maneuverability. However, the climate condition is very harsh at sea, such as typhoon, rainstorm and smog, etc. Moreover, the available space of the warship is limited, it is very difficult to erect a large antenna array. Therefore, research on the compact receiving array of HFSWR is becoming urgently.

The selection of array configuration and the pattern synthesis method of array are two important parts for the array miniaturization design. It is a target for array miniaturization design that the directivity of the system design requirement has to be satisfied when the array size is small or limited. Thus it is required that the array has super directivity.  Compared with the uniform excitation of each array element, The super directivity array has higher directive by varying the excitation mode of each array element. Compared with the common array, the super directivity array has higher directive in the same array aperture.  The directive array antenna plays an important role in designing the compact array.

At present, the optimal linear constraints [3, 4], adaptive array [5, 6] and genetic algorithm [7, 8, 9] or other intelligent optimization algorithm [10, 11] are three representative numerical pattern synthesis methods for arbitrary array. The appearance of these methods brings great help and enriches the synthesis methods of the other array form other than uniform linear array. Circular array has been paid more attention because of its ability of full plane scanning. It is a special form of planer array, it can carry out 360 degree scanning in the horizontal plane compared with linear array. The beam can uniformly scan the plane only by rotate the weight of each array element due to the rotational symmetry property. And because the array form is more compact than the linear array, thus the circular array has a special significance for the array miniaturization design.

MWSN has attracted much attention and has been used in many applications such as military target tracking and surveillance, animal state estimating and actuating, hazardous environment exploration and seismic sensing [12], which is composed of a large number of sensor nodes deployed either inside the phenomenon or very close to it [13]. A MWSN owes its name to the presence of mobile sink or sensor nodes within the network. The advantages of MWSN over static WSN are better energy efficiency, improved coverage, enhanced target tracking and superior channel capacity. In [14, 15], based on MWSN and cloud computing, the distributed parameter estimation is achieved in battlefield surveillance system. Detection and tracking of moving objects has been identified as a well-suited application, which would benefit from the use of MWSN [16-19].

In this paper, super directive circular arrays are designed and compared by using MWSN. The highest directivity with certain constrains is achieved, and with the super directive circular array as sub arrays, the size of the received antenna array is reduced. Computer simulation verifies the correctness and affectivity of the proposed method.

\section{Deployment of sensors of MWSN}
For HFSWR, the elements, which can monitor enemy aircrafts and missiles, are usually deployed near the coastline. Thus mobile nodes can be thrown down by aircraft in the deployment regions, which are far away from the military base. In order to monitor enemy air targets, deployment regions of MWSNs are vitally important to protect the military base. To the best of our knowledge, the mountainside near the coastline facing the direction of enemy attack is an excellent place to deploy mobile nodes, as shown in Fig. 1. When enemy aircrafts and missiles are coming, sensor nodes can detect the electromagnetic wave sent by enemy objects.
\begin{figure}
\includegraphics[width=3in,height=2in]{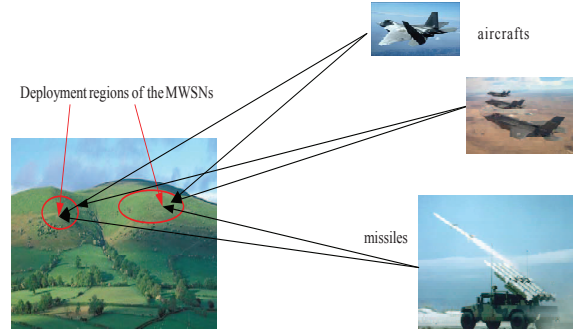}
\caption{Depolyment regions of MWSNs.}
\end{figure}

\section{Super directive circular array design}
The radiation pattern of an \emph{N} MWSN is
\begin{equation}
F(\theta ,\phi )={{w}^{H}}v(\theta ,\phi ),
\end{equation}
where $w={{[{{w}_{1}},{{w}_{2}},\cdots ,{{w}_{N}}]}^{T}}$ is the weight vector and $v(\theta ,\phi )$ is the steering vector in direction $(\theta ,\phi )$. The superscript \emph{H} and \emph{T} denote Hermitian transpose and transpose respectively. The steering vector is
\begin{equation}
v(\theta ,\phi )=\left[ \begin{matrix}
   {{g}_{1}}(\theta ,\phi )\exp (j2\pi {{f}_{0}}{{\tau }_{1}})  \\
   {{g}_{2}}(\theta ,\phi )\exp (j2\pi {{f}_{0}}{{\tau }_{2}})  \\
   \vdots   \\
   {{g}_{N}}(\theta ,\phi )\exp (j2\pi {{f}_{0}}{{\tau }_{N}})  \\
\end{matrix} \right],
\end{equation}
where ${{g}_{k}}(\theta ,\phi ),k=1,2,\cdots ,N$ are the patterns of each sensor, ${{f}_{0}}$ is the frequency and ${{\tau }_{k}}(k=1,2,\cdots ,N)$ are the propagation delays between the sensors.

The sensor directivity is defined as
\begin{equation}
D=\frac{{{\left| F({{\theta }_{0}},{{\phi }_{0}}) \right|}^{2}}}{\frac{1}{4\pi }\int_{0}^{2\pi }{\int_{0}^{\pi }{{{\left| F(\theta ,\phi ) \right|}^{2}}\sin \theta d\theta d\phi }}},
\end{equation}
where $({{\theta }_{0}},{{\phi }_{0}})$ is the looking direction. The directivity can be represented in a generalized Rayleigh quotient form using (1) and (2)
\begin{equation}
D=\frac{{{w}^{H}}Bw}{{{w}^{H}}Aw},
\end{equation}
where
\begin{equation}
A=\frac{1}{4\pi }\int_{0}^{2\pi }{\int_{0}^{\pi }{{{v}^{H}}(\theta ,\phi )v(\theta ,\phi )\sin \theta d\theta d\phi }},
\end{equation}
\begin{equation}
B={{v}^{H}}({{\theta }_{0}},{{\phi }_{0}})v({{\theta }_{0}},{{\phi }_{0}}).
\end{equation}

The directivity of the MWSN denotes the array radiation intensity averaged over all directions. Matrix \emph{A} and \emph{B} are positively definite Hermitian matrices. The maximum directivity and the corresponding weight vector can be achieved through the characteristics of generalized Rayleigh quotient [20]. The directivity synthesis problem can be transformed into a quadratic programming problem. The solution of the problem below is the corresponding weight vector.
\begin{equation}
\begin{aligned}
 \begin{array}{l@{\quad}l@{}r@{\quad}l}
 &\underset{w}{min}\,{{w}^{H}}Aw \\
 s.t.&{{v}^{H}}({{\theta }_{0}},{{\phi }_{0}})w=1 \\
 &\operatorname{Re}[{{v}_{\theta }}^{H}({{\theta }_{0}},{{\phi }_{0}})w]=0 \\
 &\operatorname{Re}[{{v}_{\phi }}^{H}({{\theta }_{0}},{{\phi }_{0}})w]=0
\end{array}
\end{aligned}
\end{equation}

The numerator of (4) is constrained to be 1 ensures that the gain in the look direction is constant. The followings are mainlobe constrains which maintain the presence of a peak in this direction. The subscripts $\theta$ ($\phi$ ) denote partial differentiation with respect to $\theta$ and $\phi$, respectively. The maximum directivity ${{D}_{max}}$ can be achieved by solving (7).

Define real vector $\tilde{w}$, $\tilde{v}$, $\hat{v}$ and real matrix $\tilde{A}$ as
\begin{equation}
\tilde{w}=\left[ \begin{matrix}
   \operatorname{Re}(w)  \\
   \operatorname{Im}(w)  \\
\end{matrix} \right],
\end{equation}
\begin{equation}
\tilde{v}=\left[ \begin{matrix}
   \operatorname{Re}(v)  \\
   \operatorname{Im}(v)  \\
\end{matrix} \right],
\end{equation}
\begin{equation}
\hat{v}=\left[ \begin{matrix}
   -\operatorname{Im}(v)  \\
   \operatorname{Re}(v)  \\
\end{matrix} \right],
\end{equation}
\begin{equation}
\tilde{A}=\left[ \begin{matrix}
   \operatorname{Re}(A) & -\operatorname{Im}(A)  \\
   \operatorname{Im}(A) & \operatorname{Re}(A)  \\
\end{matrix} \right].
\end{equation}
The ratio of external to internal noise (REIN) is defined as follows
\begin{equation}
\gamma =\frac{\int\limits_{0}^{2\pi }{\int\limits_{0}^{\pi }{\left| F{{(\theta ,\phi )}^{2}} \right|}}\sin \theta d\theta d\phi }{4\pi \sum\limits_{k=1}^{N}{{{\left| {{w}_{k}} \right|}^{2}}}}.
\end{equation}

Considering the need of anti interference and the special demands of the array pattern. An iterative procedure is adopted to make the sidelobes have uniform values equal to the desired level. Let real value function $s(\theta ,\phi )$ be the amplitude of desired response in the sidelobe region. Let ${{w}_{c}}$ denote the initial weight of each iteration, and the corresponding pattern is ${{F}_{c}}(\theta ,\phi )$. Then, the return weight of that iteration is the solution of the problem
\begin{equation}
\begin{aligned}
\begin{array}{l@{\quad}l@{}r@{\quad}l}
  & \underset{w}{\mathop{\min }}\,{{w}^{H}}Aw \\
 s.t.&{{C}^{T}}\tilde{w}=f \\
 &{{{\tilde{w}}}^{T}}\tilde{w}\le b \\
 &{{{\tilde{v}}}^{T}}({{\theta }_{i}},{{\phi }_{i}})\tilde{w}=\operatorname{Re}({{f}_{i}})\quad i=1,2,\cdots ,m \\
 &{{{\hat{v}}}^{T}}({{\theta }_{i}},{{\phi }_{i}})\tilde{w}=\operatorname{Im}({{f}_{i}})\quad i=1,2,\cdots ,m,
\end{array}
\end{aligned}
\end{equation}
where $C$ is a $2N\times M$ matrix, and $f$ is a $M\times 1$ vector. $b=1/(\varepsilon {{D}_{max}})$ , where $\varepsilon $ is the lower band of $\gamma$ and ${{D}_{max}}$ is the maximum directivity without constrains. The maximum number of controlled sidelobes ${{m}_{max}}$ is expressed as
\begin{equation}
{{m}_{max}}=\left\{ \begin{matrix}
   N-(M+1)/2\quad M\ is\ odd  \\
   N-M/2\quad \quad \ \ \ M\ is\ even  \\
\end{matrix} \right.
\end{equation}

The steps of the algorithm

(1) initialize the iteration, let ${{\varepsilon }_{0}}=\varepsilon $, giving the value of termination condition $\delta$

(2) calculate $b$ using $b=1/(\varepsilon {{D}_{max}})$

(3) set ${{w}_{c}}={{w}_{\gamma }}$, where ${{w}_{\gamma }}$ is the optimal weight vector under efficiency constrain

(4) find the \emph{m} peak of sidelobe in sidelobe region in present pattern, calculate ${{f}_{i}}(i=1,2,\cdots ,m)$

(5) calculate ${{\mu }_{0}}$ and ${{\tilde{w}}_{0}}$by solving the quadratic programming problem, calculate ${{w}_{0}}$ from  ${{\tilde{w}}_{0}}$

(6) if the array pattern satisfies the demand conditions, let ${{w}_{opt}}={{w}_{0}}$, otherwise let ${{w}_{c}}={{w}_{0}}$ and go to step (4)

(7) if $\gamma$ satisfies the termination condition, then quit, otherwise calculate ${{\varepsilon }_{0}}$ , go to step (2)

\section{Simulation Results}
First, compare uniform circular array of MWSN with even and odd number of sensors. It shows that for the MWSN with even number of sensors, the pattern varies when the direction of mainlobe changes. On contrast, for the MWSN with odd number of sensors, the patterns are almost the same for different directions of the mainlobe.

\begin{figure}
\includegraphics[width=3in,height=2in]{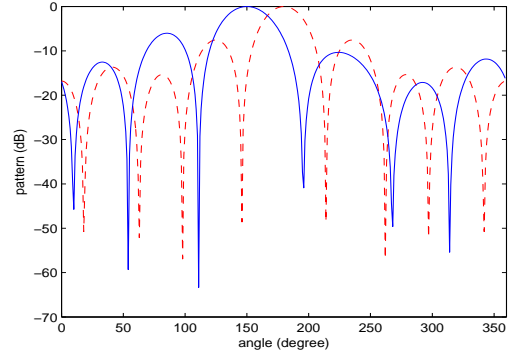}
\caption{Array pattern of 8 sensors circular array}
\end{figure}
\begin{figure}
\includegraphics[width=3in,height=2in]{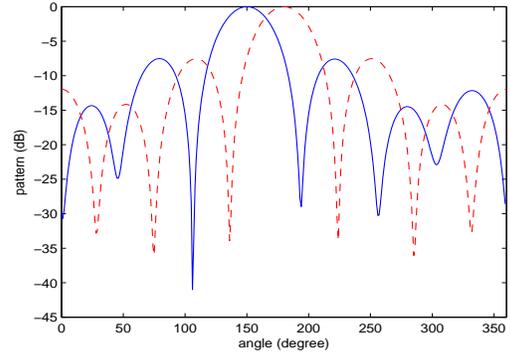}
\caption{Array pattern of 7 sensors circular array}
\end{figure}

It can be seen from Fig. 2 and Fig. 3 that it is a good choice to design super directive circular with odd number of elements. In Fig. 4, We compare the directivity and the ratio of external to internal noise of three uniform circular array of MWSN with sensor number 3, 5, 7.

\begin{figure}
\includegraphics[width=3in,height=2in]{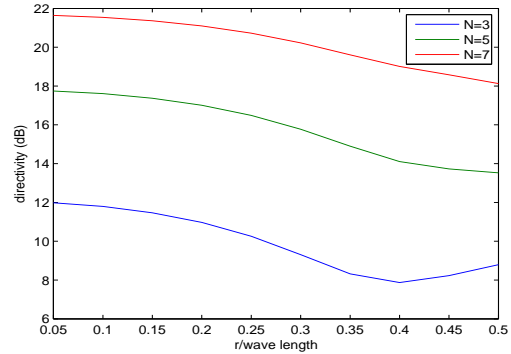}
\caption{Maximum directivity of circular arrays}
\end{figure}
\begin{figure}
\includegraphics[width=3in,height=2in]{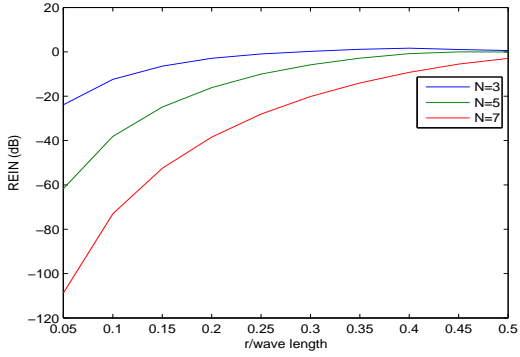}
\caption{The REIN of circular arrays}
\end{figure}

It is shown from Fig.5 that the directivity increases and the REIN decreases with the increase of the number of the array elements, the directivity decreases and the REIN increases with the increase of the radius of the array.

\begin{table}[tbp]
\centering
\caption{The requirement on the dominance of external noise in 4 to 12MHz}    
\label{tab:1}       
\begin{tabular}{|c|c|c|c|c|c|}       
\hline
f(MHz) & 4 & 6 & 8 & 10 & 12  \\ \hline
$\varepsilon $(\emph{dB}) & -30 & -23 & -19 & -16 & -13\\ \hline
\end{tabular}
\end{table}

The increase of the frequency is equivalent to the increase of the radius of the array  $(r/\lambda )$, the sensor number and radius should be determined to satisfy the REIN and sidelobe constrains from the lower band of the frequency. According to the CCIR recommendation [20], dominance of external noise condition can be ensured with the REIN $\gamma$ is no less than the $\varepsilon $ given in Table 1. At 4MHz, dominance of external noise condition is ensured with $\gamma \ge -30dB$.

It can be seen from Fig. 5 that the radius of 7 sensors array of MWSN with $\gamma =-30dB$ is $r=0.13\lambda =9.75m$, for 5 sensors array of MWSN is $r=0.053\lambda =4m$, for 3 sensor array, the radius is quite small. The required radius of 7 sensors array is too large, and the directivity is too small for 3 elements array. Thus 5 elements array should be choose.

The external noise varies on season, time and geographical location and other aspects, the constrains of the REIN should be considered, and there are special demands on sidelobes for some purpose. By using the proposed method, the weight vector can be calculated according the actual external noise, and the possibility for further reducing the array radius can be achieved.

A 5 sensors circular array with radius 3\emph{m} is used, REIN is constrained by Table 1 and sidelobe constrained -25\emph{dB}. The simulation results are shown in Table 2.

\begin{table}[tbp]
\centering
\caption{Simulation results of 5 elements array with radius 3\emph{m}}    
\label{tab:2}       
\begin{tabular}{|c|c|c|c|c|c|}       
\hline
f(MHz) & 4 & 6 & 8 & 10 & 12  \\ \hline
D(\emph{dB}) & 12.3 & 12.9 & 13.5 & 14.2 & 14.7\\ \hline
$\gamma $(\emph{dB}) & -24.0 & -17.6 & -13.4 & -10.4 & -8.0\\ \hline
\end{tabular}
\end{table}

\begin{figure}
\includegraphics[width=3in,height=2in]{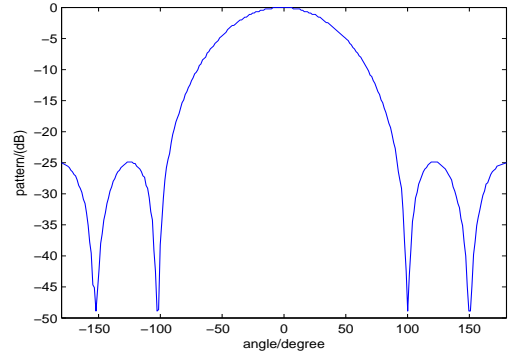}
\caption{Pattern of 5 sensors circular array}
\end{figure}
\begin{figure}
\includegraphics[width=3in,height=2in]{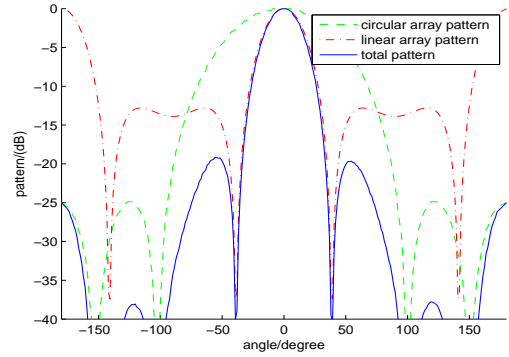}
\caption{Pattern of MWSN}
\end{figure}
\begin{figure}
\includegraphics[width=3in,height=2in]{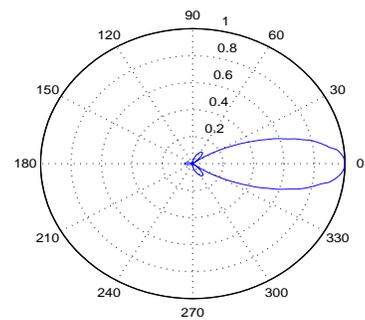}
\caption{Pattern of MWSN in polar coordinates}
\end{figure}

A receive MWSN is constructed by using 8 super directive circular arrays of sub-MWSNS as sub-arrays and placed with a distance of 15\emph{m}. Fig. 6 shows the pattern of the 5 sensors array at 4MHz, Fig. 7 shows the total array pattern, and Fig. 8 shows the total array pattern in polar coordinates. The highest directivity is achieved with sidelobe constrained under -25\emph{dB} and REIN is -24\emph{dB} satisfied the dominance of external noise condition. The total pattern is ideal, the radiation direction can be changed flexibly just by rotating the weight vector of the circular array. The array has a high directivity and a small size, which can meet the actual needs very well.

\section{Conclusion}
A super-directivity synthesis method for designing super-directivity array based on MWSN is presented. By optimizing array directivity subject to REIN and sidelobe constrains, the maximum possible directivity can be achieved. The optimized array of MWSN may have a smaller size with the same directivity compared with conventional array. By using the super directive circular array as sub-arrays, the receive MWSN can be miniaturized. The benefits of the method have been demonstrated by the simulation results. By calculating the \emph{A} and \emph{B} in (5) and (6) for other conditions, this method can also be applied to other array structures.

In the future, we will focus on direction-of-arrival (DOA) estimation in IWSNs [21-24].
\section*{Acknowledgment}

This work was supported in part by the Qing Lan Project, by the Natural Science Foundation of JiangSu Province of China, No. BK20140248, by the Educational Commission of Guangdong Province, China, Project No. 2013KJCX0131, by the Guangdong High-Tech Development Fund No. 2013B010401035, by the National Science Foundation of China under Grant 61401107.

\end{document}